**Interplay between pleiotropy and secondary selection determines rise and fall of mutators in stress response.**

Muyoung Heo and Eugene I. Shakhnovich

Department of Chemistry and Chemical Biology, Harvard University,
12 Oxford St, Cambridge, MA

Corresponding author: Eugene Shakhnovich;  Email: eugene@belok.harvard.edu

**Abstract**

The role of mutator clones, whose mutation rate is about two to three order of magnitude higher than the rate of wild-type clones, in adaptive evolution of asexual populations has been controversial. Here we address this problem by using an *ab initio* microscopic model of living cells, which combines population genetics with physically realistic presentation of protein stability and protein-protein interactions. The genome of model organisms encodes replication controlling genes (RCGs) and genes modeling the mismatch repair (MMR) complexes. The genotype-phenotype relationship posits that replication rate of an organism is proportional to protein copy numbers of RCGs in their functional form and there is a production cost penalty for protein overexpression. The mutation rate depends linearly on the concentration of homodimers of MMR proteins. By simulating multiple runs of evolution of populations under various environmental stresses – stationary phase, starvation or temperature-jump – we find that adaptation most often occurs through transient fixation of a mutator phenotype, regardless of the nature of stress, but the fixation mechanism depends on the nature of stress. In temperature jump stress, mutators take over the population due to loss of stability of MMR complexes. In contrast, in starvation and stationary phase stresses, mutators are supplied in small fraction of the population via epigenetic stochastic noise in production of MMR proteins (a pleiotropic effect), and their net supply is higher in populations of low fitness due to reduced genetic drift. Subsequently, mutators in stationary phase or starvation hitchhike to fixation with a beneficial mutation in the RCGs, (second order selection) and finally a mutation stabilizing the MMR complex arrives, returning the population to a non-mutator phenotype. Our results provide microscopic insights into the rise and fall of mutators in adapting finite asexual populations.

## Author Summary

Dramatic rise of mutators has been found to accompany adaptation of bacteria in response to many kinds of stress. Two views on the evolutionary origin of this phenomenon emerged: the pleiotropic hypothesis positing that it is a byproduct of environmental stress or other specific stress response mechanisms and the second order selection which states that mutators hitchhike to fixation with unrelated beneficial alleles. Conventional population genetics models could not fully resolve this controversy because they are based on certain assumptions about fitness landscape. Here we address this problem using a microscopic multiscale model, which couples physically realistic molecular descriptions of proteins and their interactions with population genetics of carrier organisms without assuming any *a priori* fitness landscape. We found that both pleiotropy and second order selection play a crucial role at different stages of adaptation: the supply of mutators is provided through destabilization of error correction complexes or fluctuations of production levels of prototypic mismatch repair proteins (pleiotropic effects), while rise and fixation of mutators occur when there is a sufficient supply of beneficial mutations in replication-controlling genes. This general mechanism assures a robust and reliable adaptation of organisms to unforeseen challenges. This study highlights physical principles underlying physical biological mechanisms of stress response and adaptation.

**Introduction**

Bacterial populations often respond to various stresses by inducing mutagenesis whereby mutator clones rise to fixation, at least transiently, during adaptation to stressful environments [1,2,3,4,5]. The rise of mutator clones has been observed as a universal response regardless of the nature of stress, despite the diversity of detailed molecular mechanisms associated with such responses (reviewed in [1,5]). (See, however, [6] where this interpretation is questioned for a particular experimental system.) The evolutionary significance of this observation has been controversial as two views emerged in the literature [3,7]. The pleiotropic hypothesis posits that high mutation rate is a by-product of genetic mechanisms invoked in response to stress or other physical mechanisms unrelated to adaptation [8]. The key aspect of the pleiotropic hypothesis is that high level of error correction and maintenance may be energetically costly so that bacteria would not fully activate them in stable environments [3]. Consistent with that view is the observation that natural populations exhibit a broad range of mutator allele frequencies, which are relatively higher than expected [2,9]. Higher mutation rates during adaptation may be then due to the trade-off between the requirement to repair diverse lesions in genomes and the energetic cost maintaining the fidelity of DNA polymerases involved in this process.

An alternative view is a second order selection hypothesis [10,11,12]. Mutators, which can rapidly produce beneficial mutations, could get fixed in the population by hitchhiking [12]. However, they mostly burden the population with deleterious mutations, which eventually outnumber the beneficial ones, and thus mutation rate tends to decrease to a minimum in well-adapted populations [10,13,14,15]. Computer simulations employing population genetics models provided some evidence that mutators can hitchhike to fixation when population size is large enough and stress is sufficiently profound and durable [2,15,16]. However, these theoretical studies were based on a number of phenomenological assumptions. In particular, alleles were classified into a few discrete forms such as "deleterious", "normal" and "beneficial" and fitness effect were assumed additive between alleles. Furthermore, most population genetics models are based on certain *a priori* assumptions about appearance and reversion of mutations. Implicit in these models is a peculiar effect of saturation whereby all or most alleles get fixed in their beneficial forms, essentially eliminating further supply of beneficial mutations, which causes populations to reverse to non-mutator phenotype. However, while many postulates of mathematical population genetics are rooted in experimental observations, the reality is certainly much more complex. In particular in complex crowded cellular environments, mutations in coding regions are more likely to have a broad impact on many properties of cellular proteins such as their stability, interactions with their functional and non-functional partners and of course their catalytic activity, which results in a continuous effect of mutations on fitness. Furthermore, the effect of fitness on supply of beneficial (as well as deleterious) mutations is hard to evaluate *a priori* due to enormous plasticity and size of sequence space of functional proteins [17,18].

To this end, it is very important to go beyond the phenomenological postulates of traditional population genetics models and develop a new model where population genetics is coupled to realistic yet tractable biophysical model of proteins and their interactions in cytoplasm. A first step in that direction has been made in [13] where we studied evolution of mutation rates in a population of simple organisms each carrying 3 genes. The key distinctive feature of the approach proposed in [13] is that properties of cellular proteins – their stability and interactions – were derived directly from sequences of their genomes and a simple biologically realistic relationship connected these biophysical properties with fitness (growth rate) of the model cell population.

Here, we further develop this microscopic multiscale approach to study evolutionary dynamics of stress-induced adaptation in a finite asexual population. In particular we focus on emergence (or lack thereof) of mutators during adaptation process. In the present model, each organism carries four genes expressing corresponding protein products. The first three genes are housekeeping genes responsible for cell growth and division, (replication controlling genes or RCGs), and protein products of gene 4 homodimerize to form a mismatch repair (MMR) complex – mimicking the *mut*S system in bacteria whose proteins are active in vivo as tetramers (dimers of dimers) [19,20]. While diverse molecular mechanisms can lead to stress-induced mutagenesis in bacteria (*e.g. rpo*S dependent SOS responses [21]) here we focus on a prototypical MMR system, for simplicity and because the deficiencies and down regulation of MMR genes are known in many instances to be the main cause of constitutive mutators, which are constantly supplied to the population regardless of environmental requirement, [2,9,22] as well as a major molecular event in stress-induced mutagenesis [2,23,24]. Three RCGs form the simplest functional protein-protein interaction (PPI) network where protein 1 functions in isolation and proteins 2 and 3 must form a functional heterodimeric complex. (see Figure 1)

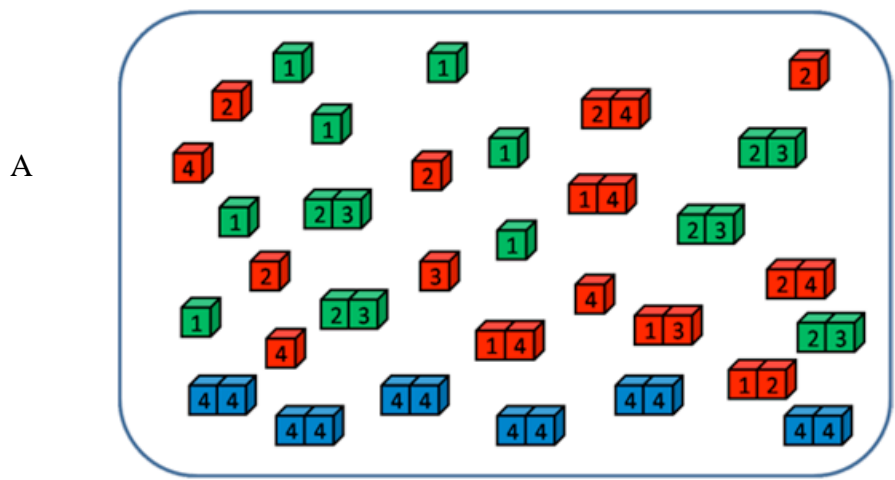

**Figure 1. A schematic diagram of the model.** A model organism has 4 genes, which are expressed into multiple copies of model proteins. Proteins can stay as monomers or form dimers whose concentrations are determined by binding constants of interactions among them and law of mass

action equations. Green cubes represent proteins in their functional states that contribute an organism's replication rate according to Eq.(1). Blue cubes represent functional MMR homodimers, whose concentration determines mutation rate of their organism. Red cubes represent proteins in their non-functional states.

The model with three RCGs was used in our recent study [13] where it was shown that this minimal model which takes into account protein function (in the form of PPI) is capable of reproducing rich biology of evolution of mutation rates. Fitness, *i.e.* the growth rate, $b$ of an organism is proportional to the monomer concentration of the protein product of the first RCG and concentration of functional dimers of protein products of the second and the third RCGs:

$$b = b_0 \frac{F_1 \cdot F_{23} \cdot P_{int}^{23} \cdot \prod_{i=1}^{3} P_{nat}^i}{1 + \alpha \left( \sum_{i=1}^{4} C_i - C_0 \right)^2}, \qquad (1)$$

where $b_0$ is a base growth rate, $F_i$ is concentration of monomeric protein $i$ and $F_{ij}$ is concentration of heterodimer complex between protein $i$ and $j$ in all possible binding configurations. $P_{int}^{23}$ is the Boltzmann probability of binding between protein *2* and *3* in the native, functional binding configuration whose binding energy has the lowest value of all possible mutual configurations and $P_{nat}^i$ is thermal tability (Boltzmann probability to be in the native state) for the protein product of gene $i$. $C_i$ is intracellular concentration for protein $i$, $C_0$ is an optimal total concentration for all proteins in a cell. Deviation from this optimal level causes drop in fitness, reflecting a metabolic cost of protein production and degradation and $\alpha$ is a control coefficient, which sets fitness penalties for deviations from an optimum production level. The importance of fitness cost for protein overproduction has been established by Dekel and Alon [25]. Phenomenologically, the overexpression cost function in the denominator of Eq.(1) prevents an artificial scenario when the increase of fitness is achieved by merely overexpressing proteins rather than by evolving their sequences.

The protein product of the fourth gene determines the mutation rate of its genome by acting as a prototype of *mut*S, which forms dimers of dimers. The fidelity of an organism's DNA replication is proportional to concentration of functional MMR homodimers formed by products of gene 4 (see Model and Methods for details). Protein concentrations $C_i$ are "inheritable" but can fluctuate, reflecting long-time correlated noise in protein production in living cells [26], and $F_i$ and $F_{ij}$ are exactly calculated for a given set of $C_i$ by solving equations of the Law of Mass Action (LMA) (see Model and Methods for details). Thus, mutation rates can increase upon a drop in concentration of functional MMR homodimers, or upon mutations of the MMR gene that disfavor its functional homodimerization, or both. (See Figure 1 and Model and Methods below for illustration and details.)

Using this *ab initio* model we study adaptation to various stresses such as higher temperature, stationary phase and starvation. In particular we focus on the importance, universality and causes of transient fixation of mutator phenotype in adapting finite asexual populations.

**Results**

Each evolutionary simulation started from a population of 500 organisms each having the same seed genome. The population size was limited at 5000 organisms so that excess organisms were randomly culled when this size limit was reached. Seed proteins sequences were optimized to have sufficiently high initial stability ($P_{nat}$) to avoid an immediate lethal phenotype (see Model and Methods for details). However neither protein sequences nor their concentrations $C$'s were optimized to achieve beneficial protein-protein interactions. Correspondingly initial fitness of the seed populations was quite low and the initial adaptation increased fitness through optimization of expression levels and protein-protein interactions (see below). Then at later time (at $t$=20000) we subjected adapted populations to stress. We modeled three types of stress. The first type was "heat shock" whereby we instantly raised temperature from $T$=0.85 to $T$=1.0 and kept it fixed afterwards. The second type of stress mimicked entrance into stationary phase whereby we instantly dropped growth rate of all organisms threefold (*i.e.* decreased $b_0$ in Eq.(1) threefold). The third type of stress simulated ''starvation'' accompanied by sharp drop in protein production. To this end we instantly dropped the optimal protein production level $C_0$ (see Eq.(1)) tenfold at $t$=20,000. For each type of stress we ran 100 simulations to obtain statistically significant results.

Fig.2 shows evolution of the populations. The first key event is an initial adaptation of seed sequences, which resulted in dramatic improvement of fitness. At this stage initial seed sequences evolve into adapted organisms where functional and non-functional PPI are optimized (see below). Three broad classes of populations (strains) distinguished by their fitness ( $b \sim 0.32, b \sim 0.62, b \sim 1$ ) emerged after initial adaptation (see Table 1 for detailed distributions), suggestive of a highly non-trivial fitness landscape in the model, containing at least three local fitness peaks. In all cases the initial adaptation occurred via transient fixation of mutator phenotype as can be seen in the bottom panel of Fig.2. Second, transient fixation of mutators took place in most cases except entrance into stationary phase in highly fit strains (Fig.2B bottom panel) which eventually did not increase fitness upon adaptation after starvation stress. For heat-shock stress the highly fit population went briefly to transient fixation of mutator phenotype but quickly eliminated it (Fig.2A bottom panel). It is also interesting to note that populations of higher fitness carried greater fraction of constitutive mutators (before stress but after initial adaptation) but after stress this relation was reversed. This is similar to experimental observation of Matic and coworkers that mutation in aging colonies is anticorrelated with the fraction of constitutive mutators [2]. The starvation stress resulted in dramatic drop of fitness for all three strains, correspondingly all three strains transiently fixed mutator phenotype upon adaptation to new conditions.

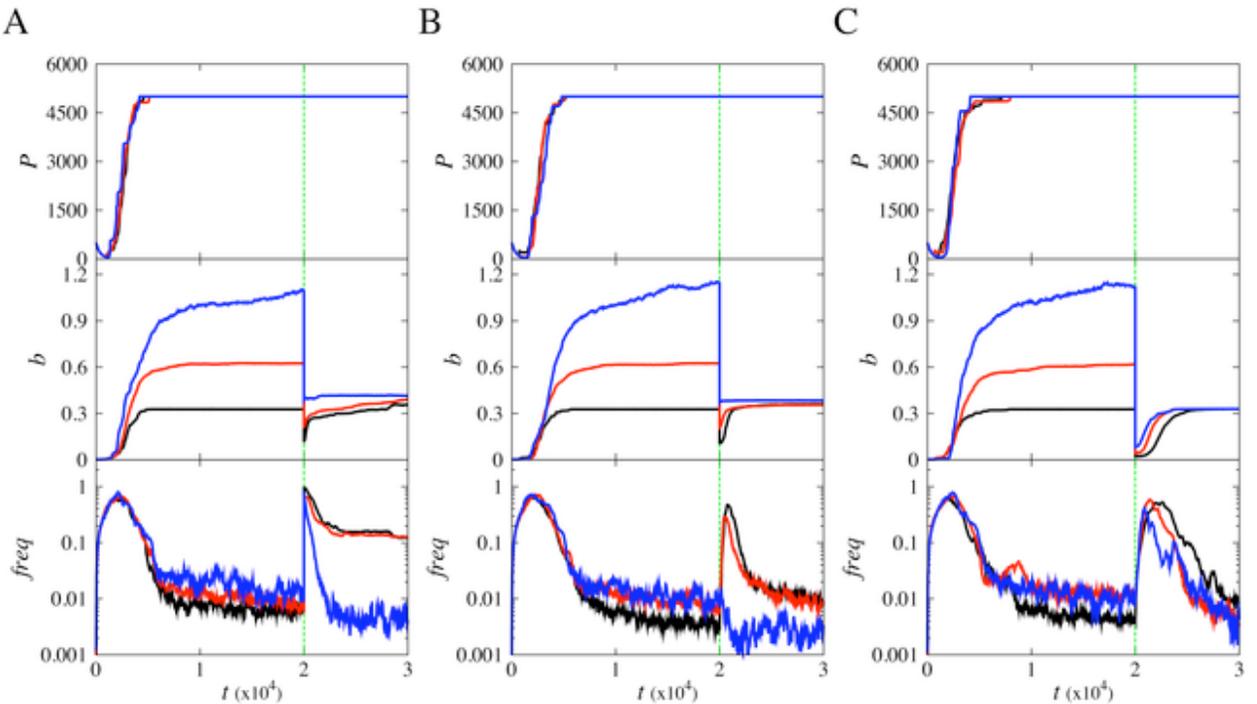

**Figure 2. Population dynamics of the model.** Panels show population ($P$), mean birth rate ($b$), and frequency of mutator allele (*freq*) in the population as function of time ($t$). All runs were classified into three groups according to fitness achieved after initial adaptation, and each curve represents averages over populations (*i.e.* evolutionary runs) within each group: $b$~0.33 (black), ~0.62 (red), ~1 (blue). Numbers of populations (runs) in each fitness group are summarized in Table I. The green lines at $t=20000$ marks the time when environmental change occurs, where (A) temperature increases, (B) the base birth rate ($b_0$) decreases by 3 fold, and (C) the optimal total concentration ($C_0$) of proteins drops by 10 fold.

Our model provides a unique opportunity to get a detailed insight into possible mechanisms, which lead to rise, fixation and fall of mutators. The dynamics of microscopic variables such as protein concentrations $C_i$, the stability of MMR proteins $P_{nat}^4$, and Boltzmann probability $P_{int}^{44}$ to form functional MMR complexes are shown in Fig.3 for the same three fitness classes (strains) of evolved populations (same color code as in Fig.2). These data provide insights into molecular mechanisms underlying the emergence, fixation and disappearance of mutator clones. Two factors are potentially responsible for the emergence of mutators: epigenetic stochastic switching through fluctuation of concentrations $C_i$ and mutations changing the stability of the MMR protein or interactions between them in a functional homodimeric complex $P_{int}^{44}$. The initial set of $C_i$ quickly converged to a more optimal distribution by reallocating resources for better fitness: The total concentration of replication-controlling proteins ($C_1, C_2$, and $C_3$) increased, while concentrations of the MMR proteins, ($C_4$) decreased. Similar parallel changes in gene expression pattern were also reported in long-term evolutionary experiments [12,23,27].

Change in concentration of the MMR protein, due to stochastic fluctuations, was the primary factor causing the rise of mutators in initial adaptation. As for adaptation to stress which took place at a later time *t*=20,000, fluctuation in MMR protein production level was primarily responsible for the rise of mutator phenotype at stationary phase and starvation stresses (see Fig.3B and Table 1), except for the strain in stationary phase stress which reached high fitness *b*~1 in initial adaptation. For this strain no further adaptation took place after stress, and mutator phenotype did not fix. For heat-shock stress the destabilization of the MMR protein and its homodimeric complex at higher temperature was the primary cause of the rise of mutator phenotype (Fig.3A). The recovery of normal, non-mutator phenotype was mostly due to mutations in the MMR gene, which increased stability of the functional MMR complex.

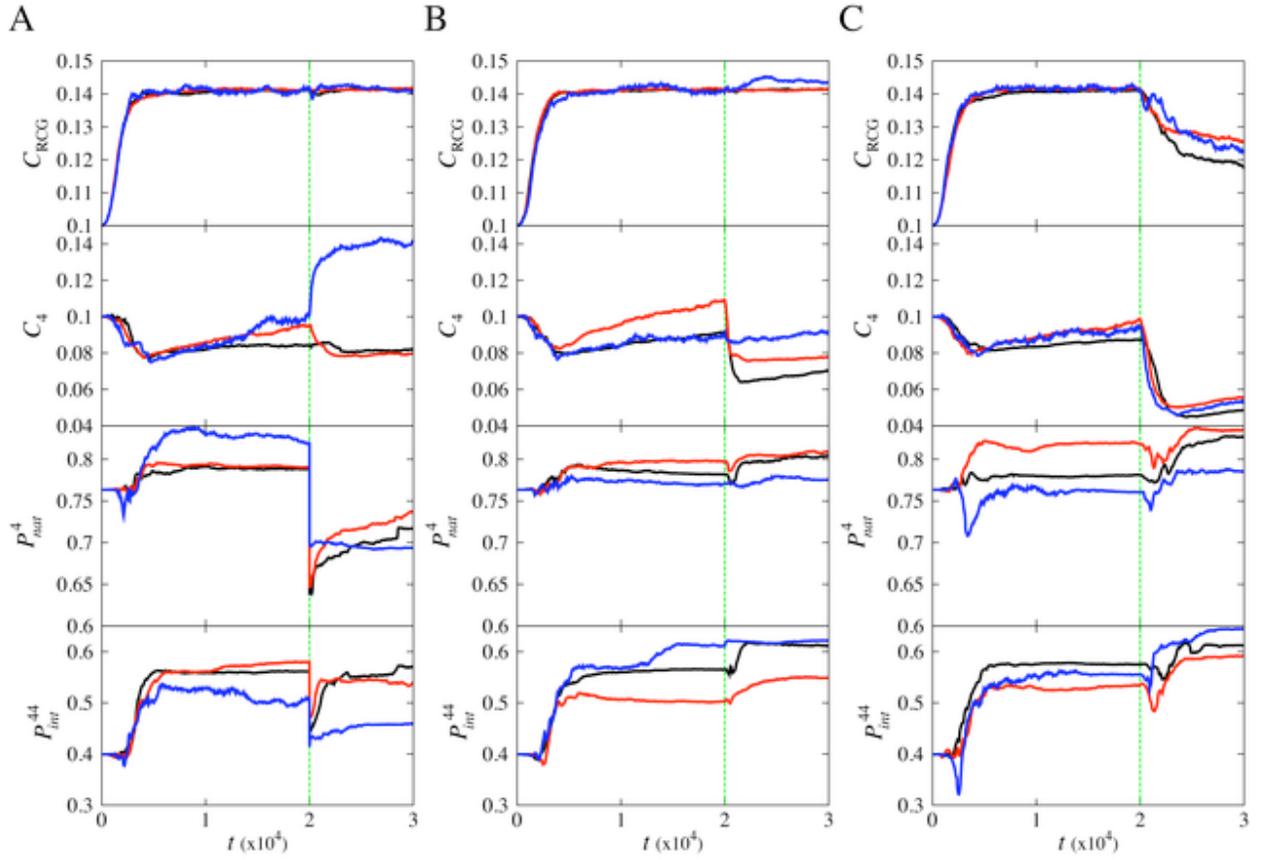

**Figure 3. Microscopic causes and consequences of adaptation events.** Panels presents mean concentrations of protein products of RCGs ($C_{RCG} = \sum_{i=1}^{3} C_i$), mean concentrations of MMR protein ($C_4$), mean protein stabilities of the MMR protein ($P_{nat}^4$) and interaction thermal probabilities of the homodimeric complex ($P_{int}^{44}$) as function of time ($t$) from top to bottom. Each column corresponds to the simulation of (A) heat shock, (B) "stationary phase" and (C) "starvation". Same grouping of populations by fitness and same meaning of line colors as in Fig.2.

In order to determine precisely the microscopic causes of phenotypic switches between mutators and non-mutators, we traced all transitions between them for all adaptation events on all trajectories. The summary picture is presented in Fig.4. Green lines in all panels of Fig.4 highlight the instances when mutator phenotype was switched on or off by variation of concentration of MMR protein $C_4$. Most mutators in the bottom panel of Fig.4, except in the temperature jump case, initially emerged from stochastic variation of protein concentrations, *i.e.* they represented switches due to epigenetic events. The transitions between fixation of mutators and non-mutators mostly occurred in a specific microscopic order, depending on the nature of stress (see Fig.4 and Table 1). The heat-shock stress resulted in thermal destabilization of the MMR complex, which gave rise to higher mutation rates. On the other hand, the stationary phase and starvation stresses decreased the growth rate which prevented the constitutive mutators from being purged away from their finite populations by genetic drift. Sequentially, highly mutating strains in all cases discovered mutations, which stabilized functional interactions in RCGs providing strains of higher fitness, so that mutator strains hitchhiked to fixation in stationary phase and starvation stresses. Finally a mutation in the MMR protein stabilized the complex bringing mutation rates in the population back to the original low level. On a microscopic level, the behavior of generating mutator strains in response to temperature stress is somewhat different from the behavior to stationary phase and starvation stresses. In the former case stress induces mutator strain directly by disrupting the MMR complex, while in the latter case it does not induce mutator strains *per se* but set in motion a chain of microscopic and populational events, such as hitchhiking, which result in a similar phenotypic phenomenology as adaptation to temperature jump.

| ES | Tran. | fitness | pathway | env | | mut | | SS | |
|---|---|---|---|---|---|---|---|---|---|
| **Tj** | M | ~0.33 | 27 | 26 | (96.3%) | 0 | (0.0%) | 0 | (0.0%) |
| | | ~0.66 | 54 | 44 | (81.5%) | 0 | (0.0%) | 5 | (9.3%) |
| | | ~1 | 10 | 5 | (50.0%) | 0 | (0.0%) | 0 | (0.0%) |
| | W | ~0.33 | 26 | - | - | 18 | (69.2%) | 4 | (15.4%) |
| | | ~0.66 | 49 | - | - | 33 | (67.3%) | 10 | (20.4%) |
| | | ~1 | 5 | - | - | 3 | (60.0%) | 2 | (40.0%) |
| **Sp** | M | ~0.33 | 35 | 0 | (0.0%) | 4 | (11.4%) | 23 | (65.7%) |
| | | ~0.66 | 42 | 0 | (0.0%) | 4 | (9.5%) | 9 | (21.4%) |
| | | ~1 | 16 | 0 | (0.0%) | 0 | (0.0%) | 0 | (0.0%) |
| | W | ~0.33 | 27 | - | - | 23 | (85.2%) | 4 | (14.8%) |
| | | ~0.66 | 13 | - | - | 10 | (76.9%) | 3 | (23.1%) |
| | | ~1 | 0 | - | - | 0 | (0.0%) | 0 | (0.0%) |
| **St** | M | ~0.33 | 49 | 0 | (0.0%) | 8 | (16.3%) | 40 | (81.6%) |
| | | ~0.66 | 33 | 0 | (0.0%) | 3 | (9.1%) | 30 | (90.9%) |
| | | ~1 | 11 | 0 | (0.0%) | 1 | (9.1%) | 7 | (63.6%) |
| | W | ~0.33 | 48 | - | - | 35 | (72.9%) | 13 | (27.1%) |
| | | ~0.66 | 33 | - | - | 27 | (81.8%) | 6 | (18.2%) |
| | | ~1 | 8 | - | - | 6 | (75.0%) | 2 | (25.0%) |

**Table 1. Causes of transitions into mutator (M) and non-mutator (W) clones: an environmental change (env), a mutation (mut) and epigenetic stochastic switching (SS) due to the noise of protein production level.** We performed 100 independent runs for populations, which respectively experience environmental temperature jump (Tj), stationary phase (Sp) and starvation (St) as environmental stress (ES). Different runs achieve different levels of fitness after the initial adaptation. They can be broadly classified into three groups to which most of the runs belong - which are ~0.33, ~0.62, and ~1, and third column shows the number of runs which reached a corresponding fitness level. Columns 4-6 give numbers of runs where rise and fall of mutators was attributed to a specific molecular cause during the second adaptation events after $t$=20000.

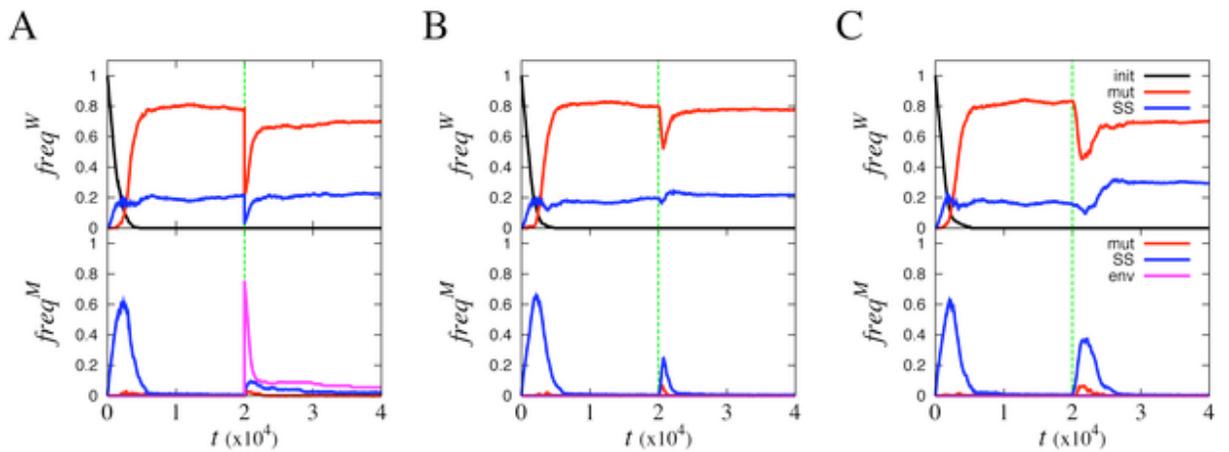

**Figure 4. Causes of rise and fall of mutator clones.** Frequencies of non-mutator ($freq^W$) and mutator ($freq^M$) clones that arise from various causes are shown as function of time ($t$). Colors indicate main molecular causes of rise and fall of mutators: Red represents genotypic mutation (mut), blue represents stochastic phenotype switching (SS), and magenta represents an environmental change (env). The black line is the original non-mutator population. Panels A, B, and C respectively correspond to population dynamics in undergoing heat shock, stationary phase, and starvation stress at $t=20000$.

Why did mutators preferentially emerge through epigenetic stochastic switching rather than a genotypic change (mutation)? To address this question we studied adaptation in response to the stress of stationary phase at various rates $r$ of stochastic fluctuation of protein concentrations, from $r=10^{-2}$ to $10^{-3}$, $10^{-4}$, and $r=0$ – the case where no fluctuations of protein concentration were allowed (Fig. 5; see Model and Methods and Table 2 for definition of fluctuation rates $r$). To make a right comparison among simulations with four different conditions, we assigned unequal concentrations to the RCG proteins and MMR protein setting them similar to those reached after the first adaptation event, otherwise the inability to relax an imbalance among equally fixed protein concentrations at the control of $r=0$ might constraint the evolution of fitness.

Deceleration of fluctuation rate delayed fixation of mutators, and furthermore, no mutators (and, strikingly, adaptation) were observed when $r=0$.

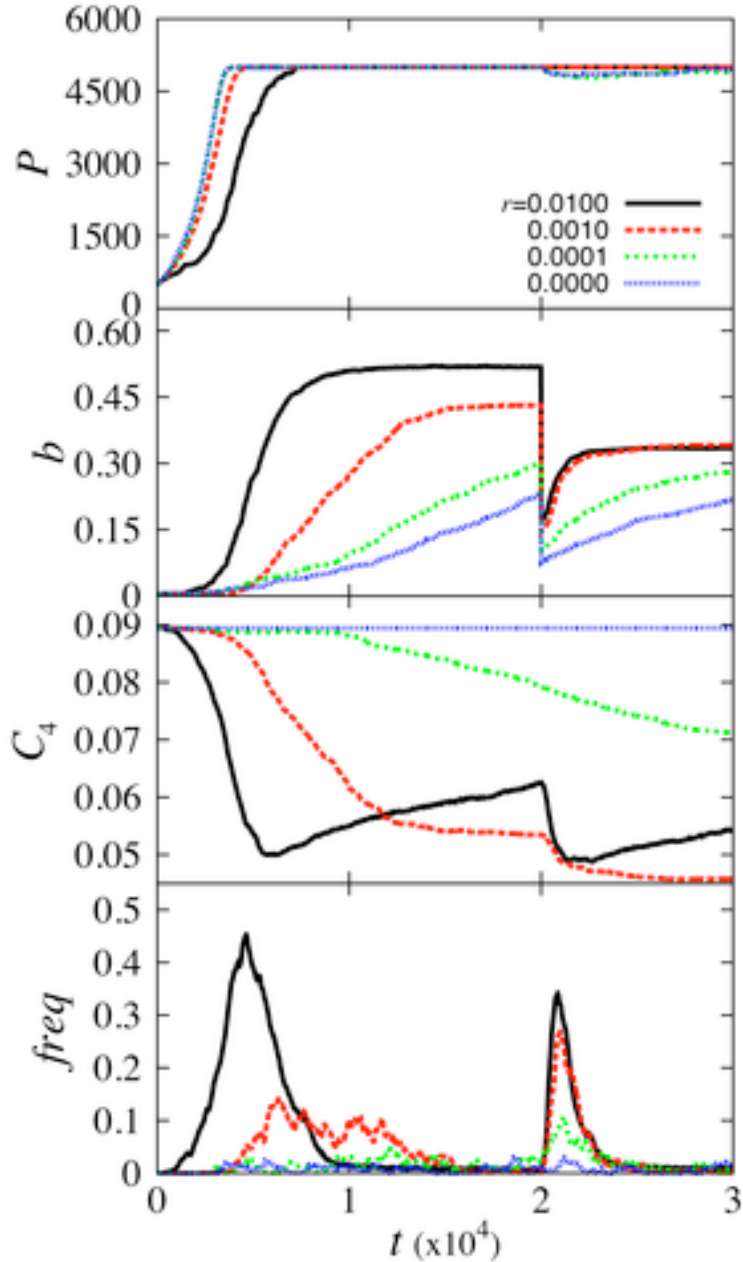

**Figure 5. The importance of stochastic switching for mutator fixation and adaptation.** Population ($P$), mean birth rate ($b$), mean total concentration of MMR protein 4 ($C_4$), and frequency of mutator allele (*freq*) in the population are plotted as function of time ($t$). The lines represent simulations at various expression level fluctuation rates: $r=10^{-2}$ (black), $10^{-3}$ (red), $10^{-4}$

(green) and 0 (blue). The initial growth of the population with higher fluctuation rate is limited due to the genetic load of deleterious mutation caused by high frequencies of mutators (see the upper panel and Fig.6). All traces for *r*>0 showed decreased gene expression levels ($C_4$) of MMR protein during and immediately after adaptation events. The concentrations at *r*=0.01 (black curves) decayed fast due to high fluctuation probability, while the decays of concentrations at lower *r* (red and green) appeared less and slower. Without stochastic switching (blue), no fixation of mutators could arise and adaptation was severely impaired.

The upper panel of Fig.5 points out to a peculiar feature: while populations with highest rate *r* of protein copy number fluctuation evolved to highest fitness *b*, their initial population growth was not the highest. In order to resolve this apparent contradiction we carried out a simulation where species with high rate of protein concentration fluctuations *r*=0.01 competed with the ones with no protein concentration fluctuations. (Fig.6). For the first 500 time steps, the fractional population of highly fluctuating *r*=0.01 species decreased. The species with high fluctuation rate provided more mutators due to epigenetic stochastic switching and their high mutation rate effectively reduced the growth rate of the population due to the heavy genetic load of deleterious mutations. Fitness curves shown in Fig.6 (red and blue curves) indicate that initial drop in fractional population of the *r*=0.01 species (black curve) was not caused by the difference in growth rates between two competing species. The initial decrease of fractional population of the *r*=0.01 species is reversed after it found a beneficial mutation in RCGs which provided higher fitness at t~500 and the population with *r*=0.01 started to dominate in the competition. We conclude that the genetic load of high mutation rate initially burdens the population with *r*=0.01, which is enriched in mutators and its growth curve is effectively limited until it finds a beneficial mutation.

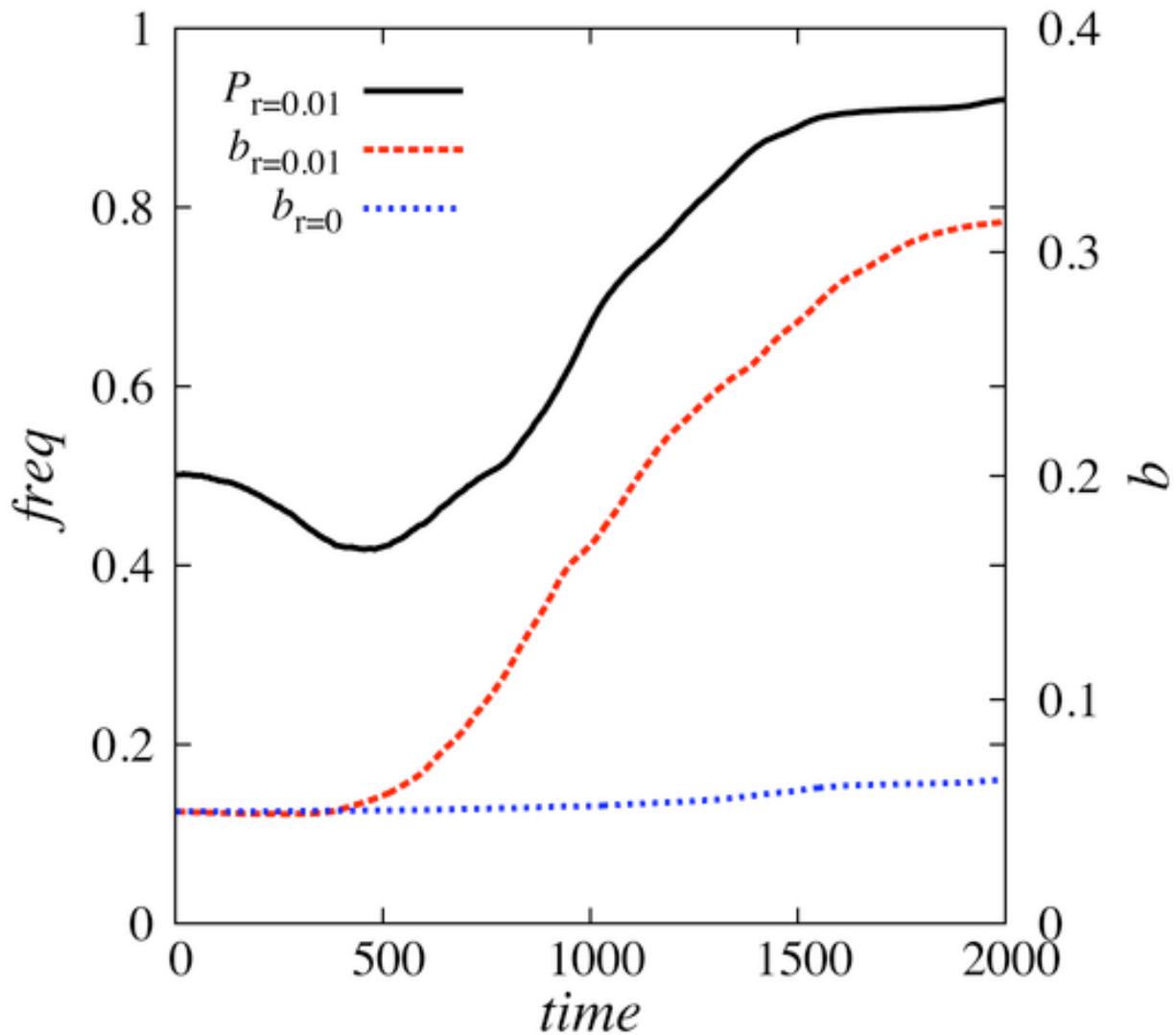

**Figure 6. The genetic load of deleterious mutations.** Simulations of competition (averaged over 100 runs) between populations with fluctuating concentration of proteins ($r=0.01$) and populations where protein concentrations are kept fixed ($r=0$). We initially seeded 500 organisms each for $r=0.01$ and $r=0$, and measured the fractional population (*freq*) of species with high fluctuation probability $r=0.01$ (black curve). It decreased for the first 500 time steps. Species with high fluctuation probability provided more mutator phenotypes due to the stochastic switching and high mutation rate effectively reducing the growth rate of the population. Fitness (*b*) of two competing populations, drawn in red dotted curve for $r=0.01$, and in blue dotted curve for $r=0$, shows that initial decrease of fraction of organisms with $r=0.01$ does not result from their growth rate difference. A decreased fraction of fluctuating ($r=0.01$) phenotypes in the population is restored after they found a beneficial mutation in RCGs which promote fitness at $t\sim500$ and the fluctuating population started to dominate in the competition.

Visser *et al.* showed that the initial level of fitness of the founder population dramatically affects the rate of adaptive evolution: the rate of adaptation was much slower for populations founded by an adapted strain than for the populations founded by an initially unadapted strain [28]. This finding is in direct agreement with our results. Fig.2 shows that populations that achieved high fitness ($b=1$, blue lines) did not further evolve after stationary phase stress and only briefly fixed mutators upon heat shock with no significant adaptation afterwards, while in starvation stress where fitness dropped more significantly all three strains showed some degree of adaptation (see Fig.2C). In contrast, less evolved populations adapted significantly after stress by reaching the characteristic level of fitness of more adapted populations at longer times. Most importantly, such difference in post-stress adaptation patterns is directly matched by the difference of the frequencies of mutators caused by stress: it is markedly narrower for initially well-adapted population than for less adapted populations (see Fig.2). Visser *et al.* hypothesized that more adapted populations have lower supply of strongly beneficial mutations, making the wait time for them to arrive longer [28]. In order to evaluate the importance of changes in fitness landscape upon adaptation we determined local "fitness landscapes" of populations (*i.e.* distributions of relative fitness change upon point mutations), immediately prior to heat shock, after heat shock and after post-stress adaptation (at $t=25,000$) (see Fig.7). We found that while differences in fitness landscapes may be noticeable, there were no pronounced patterns of differences except for extremely rare mutations that change fitness significantly which were not found in populations, which adapted to high temperature. Furthermore, our simulations also confirmed that the mutators were able to get fixed in the populations in response to the stresses of stationary phase and starvation, because the overall patterns of fitness landscapes were conserved against those stresses.

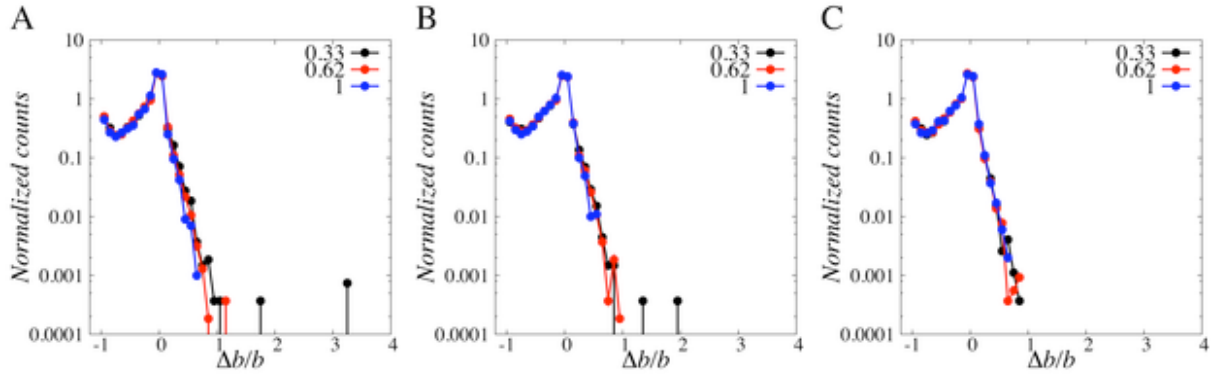

**Figure 7. Local fitness landscapes: Distributions of relative fitness changes upon a single mutation.** In order to evaluate changing fitness landscapes of different populations, we selected 100 organisms, one from each run, which represent 100 independent populations at certain time points (A) of Fig.2. The organisms' genomes were then each subjected to 1000 separate non-synonymous random point mutations and the birth rate was calculated for each of 1000 mutants per representative organism according to Eq.(1) assuming that total protein concentrations $C$ are unchanged by point mutations. $\Delta b/b$ gives the relative change in fitness upon a point mutation. The histogram counts averaged number of mutants at a certain $\Delta b/b$. All runs were classified into three fitness groups as before and each group is color-coded in the same way as explained in the caption of Fig.2. Each plot shows the fitness landscape of organisms (averaged within each fitness group) at a certain time point in the simulation: (A) $t=20000$ and $T=0.85$, before temperature increase in the heat shock adaptation simulation, (B) $t=20000$ and $T=1.00$, immediately after temperature increase, and (C) $t=25000$ and $T=1.00$, after adaptation to higher temperature.

What is then an explanation for the bias to emerge and fix mutators in lesser adapted strains? In order to address this question we performed a control simulation where fitness is constant, independent on sequences (*i.e.* not determined by Eq. (1)), so that supply and fixation of mutators are decoupled. Since the main reason for *fixation* of mutators in case of "stationary phase" and "starvation" stresses appears to be hitchhiking with beneficial mutations in RCG, by eliminating hitchhiking sequence-independent fitness model focuses entirely on supply of mutators rather than their fixation. We compared average fraction of mutators in populations having different values of fixed fitness *b*. However, we still left a protein structural constraint which removes organisms due to protein malfunction if any of its proteins lost stability *i.e.* its $P_{nat} < 0.6$. This constraint provided a weak selection against deleterious mutations, which arose more frequently in mutator clones. The results shown in Fig.8 suggest that populations of higher fitness contain less mutators. In order to understand this finding, we note that our chemostat regime simulates a limited-resource environment in which excess organisms are removed at random. High fitness in such an environment causes greater production of new organisms and hence a larger excess of organisms over the carrying capacity. Thus, more organisms must be culled at high fitness per unit time, which means a faster random drift. The trend shown in Fig.8 indicates that the level of fitness determines the frequency of mutator clones through random drift, because mutators are supplied at a constant rate by epigenetic stochastic switching and fitness determines the rate at which they are purged from the population due to genetic drift.

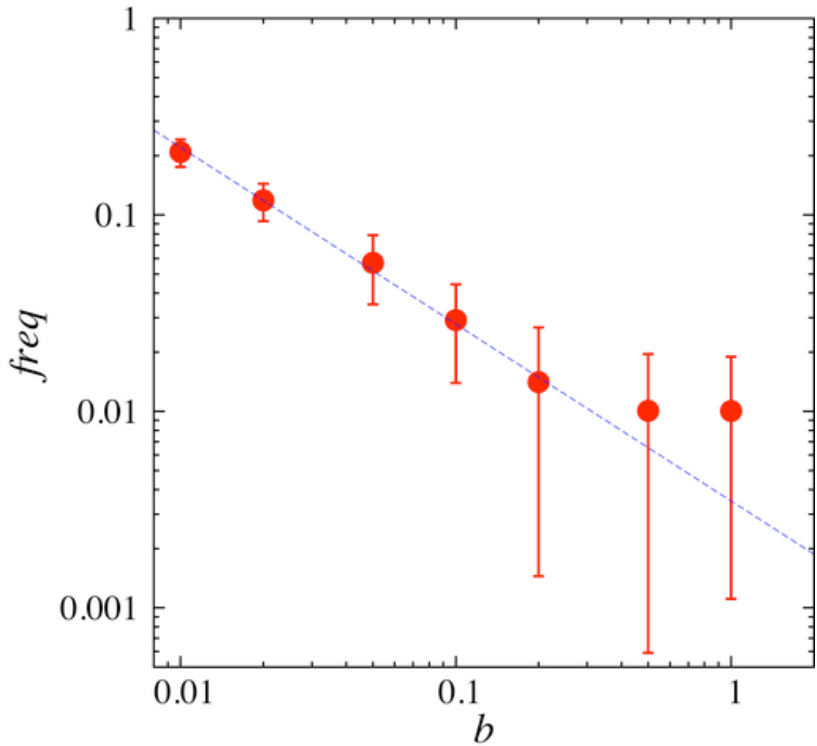

**Figure 8. Analysis of random drift through control simulations with constant fitness.** Fitness (*b*) vs. the fractional population of mutators (*freq*) is plotted on a log-log scale. Each point represents the fractional population of mutators at different constant birth rates. We performed 50 independent simulations for each condition and sampled the fractional population of mutators every 25 time steps from $t=8000$ to $t=10000$ (total 80 steps per each simulation) to obtain the ensemble averaged data. The blue dashed guideline is also plotted as a function of $0.0035 \cdot b^{-0.9}$. Because the constant fitness condition neutralized the effect of mutations on fitness, the fraction of mutators in a population did not depend on hitchhiking and was completely determined by the amount of random genetic drift that corresponds to each level of fitness. Stochastic phenotypic switching caused by noise in protein production levels continuously supplied mutators to the population, but higher growth rate increased the speed of random genetic drift. Thus, mutator clones were more rapidly eliminated at higher fitness, so that mutator frequency dropped as growth rate increased.

**Discussion. Universality and diversity of stress responses**

In this work we presented a model, which combines biophysical principles of protein folding and protein-protein interactions in crowded cellular environments, with population genetics and applied it to study universal principles of adaptation in asexual populations. The model is still mesoscopic as it includes simplified representation of proteins and their functional and non-functional interactions. However it is much more detailed and microscopic than more traditional population genetics models of evolution of mutation rates [14,15,16,29,30] because it derives fitness directly from an organismal genotype and protein concentrations in the cell and therefore can directly and explicitly assess the evolutionary consequences of genomic mutations. Unlike the conventional population genetics models, our model does not make *a priori* assumptions about fitness effect and supply of beneficial and deleterious mutations and it does not assume a fixed fitness landscape or for that matter any *a priori* fitness landscape. Rather, this multi-scale model is based on a number of intuitive biological assumptions. First, in order to function, proteins have to be in their native (folded) state and participate in functional protein-protein interactions, when needed. Proteins in this model (and of course in reality) may participate in non-functional interactions (red boxes in Fig.1), however that would result in lower copy numbers of proteins available for functional interactions and biological activity and consequently would lead to lower fitness of an organism. Second, our model considers two types of genes: housekeeping genes (or RCGs) and genes that carry out control/regulation function, in this case gene 4, whose product is responsible for control of mutation rates. Fitness of an organism is proportional to concentrations of replication controlling proteins *in their functional form* as stipulated by Eq. (1). Third, production of proteins incurs a cost invoking a fitness penalty for overproduction. This constraint makes it detrimental for the total concentration of all proteins to go beyond some optimal level, and therefore it causes in some cases redistribution of resources between productions of different proteins rather than the increase of overall protein production. Expression levels of different genes determining, along with other factors, copy numbers/concentrations of their protein products can fluctuate on time scales, which are much faster than time scales of mutations in upstream regions which also cause changes in protein productions and be "inherited", reflecting epigenetic phenomena. This factor reflects extrinsic noise in gene expression, which were observed in many cell types [31,32]. The "inheritance" of protein concentrations is not a genetic phenomenon - rather it is due to long-time correlations in extrinsic noise in protein production which were found by Elowitz and coauthors [26]. Fourth, model cells replicate at the rates corresponding to their fitness levels, so that their population can grow until it reaches a threshold size, after which excess organisms are removed randomly to maintain fixed population size. This process sets an effective total death rate, which is equal to replication rate when population size is kept fixed. The prototypic system to model mutational control here is a mismatch repair (MMR) system, which involves several proteins that are functional in their dimeric (*mut*L) or tetrameric (dimer of dimers) from (*mut*S) [20,33] Accordingly our model MMR proteins are functional in a homodimeric form. While diverse molecular mechanisms exist which determine mutation control under known stressful environments (*e.g. rpo*S dependent responses to DNA damage and other known stress responses [1,5,21], most of the mutator bacteria isolated in the laboratory and in nature have been shown to downregulate or be defective in the MMR system [2,10,22]. Most importantly, changes in expression level or mutations in MMR system proteins represent most universal response to stress regardless of the origin of bacteria or the character of challenge. Our aim here is to elucidate the role of mutation rates in stress response in an *ab initio* model and therefore using the MMR system as a prototype appears to be a logical choice.

In this study we investigated three types of stresses, which affect different properties of our model cells. At temperature stress mutator phenotype emerges simultaneously in most organisms due to destabilization of MMR complexes at higher temperature. There is no follow-up mutator fixation stage in this case (Fig.2). The rise of mutators after temperature jump is a clear example of a pleiotropic phenomenon where physical factors rather than adaptive mechanisms are responsible for the rise of mutators in the population. Second order selection does not play a significant role in the rise of mutators in temperature jump but plays a role in their fall by providing stabilizing mutations in the MMR complex, which bring mutation rates in the population back to the original low level. The sequence of events upon adaptation in two other types of stresses is quite different from the temperature jump stress. Here, both pleiotropy and second order selection play an important role in rise of mutators. Both decrease of the baseline value $b_0$ or drop in the optimal protein production level $C_0$ lead to an instant drop of fitness for all organisms (see Fig.2). Why would then such a uniform change as drop in $b_0$ result in a response? It may seem, at a first glance, that drop in $b_0$ should be equivalent to change in time scales without any material consequences. However, we found that the immediate consequence of fitness drop is the increased supply of mutators in the population of fixed size due to diminished genetic drift (Fig.8). The reason for that is the interplay of two time scales: a faster time scale at which fluctuations in protein production level supply organisms with mutators whereby MMR complexes fail to dimerize, and the time scale at which excess organisms are randomly killed in the environment which maintains a finite population size. As a result at lower fitness level the *net* supply of mutators is greater providing a necessary diversity of mutation rates in the population which will give rise to subsequent fixation of mutators via hitchhiking. The initial supply of mutators is certainly a pleiotropic phenomenon in the sense that it is caused by physical processes, which are unrelated to adaptation. Now increased supply of organisms which have higher mutation rates provides ample opportunity to acquire mutations in RCGs, which increase fitness of an organism. This is clear from Supplementary Figure 1, which provides clear evidence that mutations in RCGs are responsible for all increases in fitness. Fixation of mutator phenotype in this case is a classical example of hitchhiking, *i.e.* second order selection.

Our model points out at noise in protein production levels as a major epigenetic pleiotropic source of mutators in stationary phase and starvation adaptation. The key feature of this mechanism is that it epigenetically produces greater diversity of mutation rates in populations than would have been possible due to genotypic diversity only at a very low natural mutation rate of approximately 0.003 mutations per genome per generation [34]. This factor supplies mutators, which improve fitness through beneficial mutations in RCGs (see Supplementary Figure 1). Other hypothetical possibilities such as an elevated mutation rate in the upstream regions of the MMR genes might generate similar diversity, however we do not have evidence that such mechanism does indeed exist.

In real biological systems noise-induced mechanism, which supplies mutators, can be supplemented and strengthened by directed regulation of copy numbers of MMR proteins. Experiments show that expression of *mut*S or *mut*L genes are often downregulated upon entering into stationary phase [23,24,35]. A decrease in copy number of MMR proteins, is predicted by our model as a universal initial step in adaptation in stationary phase and starvation leading to a quick transient fixation of mutator clones. A key component of the *Escherichia coli* MMR system, *mut*S, is efficient in its tetrameric form as dimer of dimers. It is noteworthy that at the conditions of exponential growth, concentration of *mut*S dimers is close to the threshold of the dimer-tetramer equilibrium transition [23,36]. The proximity of the concentration of the MMR components to the critical threshold makes the number of functional *mut*S tetramers most susceptible to noise and it explains the persistent presence of a small proportion (1-10%) of mutators in the adapted populations observed in our simulations (Fig.2) and in experiment [10,27]. The importance of noise in gene expression for adaptation is indirectly supported by the observation that expression of stress-related proteins in *Saccharomyces cerevisiae* is controlled by TATA-containing promoters which are known to give rise to noisy gene expression while housekeeping genes are mainly under TATA-less promoters [37,38]. Furthermore, Blake and coauthors showed that increasing noise in expression of stress related genes (by mutating the TATA region) resulted in greater benefit in adaptation to acute environmental stress [39]. More immediately, we predict that modulating noise in production of *mut*S proteins in *E. coli* without affecting the mean (*e.g.* by introducing mutations which decrease binding affinity of dimers concurrently increasing the expression level of the gene) would result in dramatically altered response to an unknown stress. The work to test these predictions is underway.

Our study highlights an important interplay of pleiotropic and genetic factors in generating mutator clones and suppressing them when population adapts. In particular Table 1 shows that the dominant mechanism by which populations return to normal mutation rates after adaptation is genetic - acquiring a mutation in the MMR gene which makes the complex more viable. The important role of recurrent losses and reacquisition of MMR gene functions was highlighted in the study by Denamur *et al.* who found that phylogeny of the MMR genes in *E. coli* is very different from that of the housekeeping genes [40]. These authors found the evidence that horizontal gene transfer of MMR genes may play an important role by increasing the rates of reacquisition of MMR function over those expected from compensating mutations only as implemented in our model. While our model does not allow for gene transfer it also points out an importance of changes in MMR genes in adapting populations. Gene transfer mechanisms may make these processes faster eliminating the need to wait for a specific point mutations in the MMR genes.

Our microscopic evolutionary model of mutations and adaptation in populations of asexual organisms is still simple and minimalistic. It represents proteins at a coarse-grained level. Another important simplification of the model is mean-field treatment of PPI using the LMA approach. Such approach is good at time and length scales at which a protein participates, permanently or transiently, in multiple PPI. While certainly applicable to highly expressed proteins, the LMA treatment may be an oversimplification for proteins whose copy numbers in a cell is small. In this case either direct simulation of PPI in crowded cellular environment as in [41] or corrections to mean-field LMA as in [42] would be required for a more complete analysis. Furthermore, our simple 4-gene model despite its explicit character, is certainly a major simplification of reality with its thousands of genes operating in a crowded cytoplasm. Nevertheless, the unique feature of this approach, in contrast to traditional population genetics studies of mutation rates, is that it couples first principles consideration of protein folding and protein-protein interactions with population genetics. We find that important aspects of our findings are due to the fact that fitness landscape is not *a priori* pre-determined but is evolving as populations evolve. As such, this model provides a description of physical principles of adaptation on all scales, from individual proteins to their assemblies in cytoplasm to populations of asexual organisms. On the population level, we found that adaptation always proceeds through transient fixation of a mutator phenotype (except in cases of high fitness pre-stress populations).

This is realized, on a microscopic level of proteins and their interactions, through a sequence of events, which involve a peculiar interplay of intrinsic noise and genomic variation. Utilization of noise in protein copy numbers to trigger a set of adaptation events provides a clear evolutionary advantage in meeting unforeseen challenges for which no detailed molecular response mechanism may be available. The fact that a minimalistic "first principles" model was able to describe realistically many principal aspects of molecular and cellular mechanisms of adaptation in real bacteria suggests that evolution uses general physics as its "design scaffold", around which it builds a beautiful structure of living cells.

## Model

In our model, organisms carry 4 genes whose sequences and structures are explicitly represented. Each gene contains 81 nucleic acids, encoding 27-mer model proteins. Once it is expressed into a protein, it folds into a 3x3x3 compact lattice structure [43,44]. Lattice models have been instrumental in gaining key insights into mechanisms of protein folding [44,45,46,47], protein design [17,48] and evolution [49,50,51].

We reduce the range of all possible 3x3x3 lattice structures, which totals 103,346 [43], to randomly chosen and evenly distributed representative set of 10,000 structures for faster calculation. $P_{nat}$ is the Boltzmann probability that a protein stays in its native structure whose energy is the lowest out of all 10,000 structures. There exist 144 rigid docking modes between two 3x3x3 lattice proteins, considering 6 surfaces for each protein and 4 rotations for each surface pair of two proteins (6x6x4). $P_{int}^{ij}$ is the probability that two proteins $i$ and $j$ form a stable dimeric complex in the correct docking mode. $P_{nat}^{i}$ and $P_{int}^{ij}$ are proportional to the Boltzmann weight factors of the native structure energy, $E_0$, and the lowest binding energy, $E_0^{ij}$ as follows:

$$P_{nat}^{i} = \frac{\exp\left[-E_0^i/T\right]}{\sum_{k=1}^{1000}\exp\left[-E_k^i/T\right]}; \quad P_{int}^{ij} = \frac{\exp\left[-E_0^{ij}/T\right]}{\sum_{k=1}^{144}\exp\left[-E_k^{ij}/T\right]}. \quad (2)$$

Where $E_0^i$ is energy of the native, *i.e.* the lowest energy, conformation (out of all 10,000 conformations) of a protein, which is a product of gene number $i$ $(i=1,2,\cdots,4)$. $E_0^{ij}$ is energy of native, *i.e.* the lowest energy, binding mode (out of 144 possible ones) between proteins, which are products of genes $i$ and $j$ $(i = 1,2,\cdots,4, j = 1,2,\cdots,4)$.

The binding constants $K_{ij}$ between proteins *i* and *j* are calculated as follows:

$$K_{ij} = \frac{1}{\sum_{k=1}^{144} \exp\left[-E_k^{ij}/T\right]}, \qquad (3)$$

and these values are substituted into the Law of Mass Action (LMA) equations in Eqs. 5 and 6 to determine free concentrations of proteins $F_i$ and concentrations of their complexes $F_{ij}$. We use Miyazawa-Jernigan pairwise contact potential for both protein structural and interaction energies [52]. We report environmental temperature *T* in Miyazawa-Jernigan potential dimensionless energy units.

Simulations start from a population of 500 identical organisms (cells) each carrying 4 genes with initial sequences designed to be stable in their (randomly chosen) native conformations with $P_{nat} > 0.6$. At each time step, a cell can divide with probability *b* given by Eq.(1). A division produces two daughter cells, whose genomes are identical to that of mother cells apart from mutations that occur upon replication at the rate of *m* mutations per gene per replication. Mutation rate *m* depends on concentration of functional (homodimeric) MMR proteins (products of gene 4) as specified below. The stability loss of any protein by a mutation ($P_{nat} < 0.6$) incurs lethal phenotype [53], and the cell carrying such gene is discarded. Constant death rate, *d*, of cells is fixed to 0.005/time unit, and the parameter $b_0$ is adjusted to set the initial birth rate equal to the fixed death rate (*b=d*). The control coefficient $\alpha$ in Eq. (1) is set to 100. All parameters are listed in Table 2.

| parameter | description | value |
|---|---|---|
| $b_0$ | Base growth rate in Eq. (1) | 707.445 |
| $b_0'$ | Base growth rate modified from $b_0$ to simulate stationary phase | 235.815 |
| $d$ | Constant death rate of cells | 0.005 |
| $r$ | The rate of protein expression level fluctuation | 0.01 |
| $\alpha$ | Protein production level constraints coefficient | 100 |
| $C_0$ | Optimal production level for proteins in a cell | 0.4 |
| $C_0'$ | Optimal production level for proteins in the cell under a starving environment | 0.04 |
| $m_0$ | Base mutation rate in Eq. (2) | 0.0001 |
| $T$ | Environmental temperature in an arbitrary unit | 0.85 |
| $T'$ | The high environmental temperature achieved by T-jump simulation. | 1.00 |

**Table 2. Simulation parameters.** The parameters used in the simulation are listed in this table.

We simulated a chemostat regime: when the population size exceeded 5000 organisms, the excess organisms were randomly culled to bring the total population size back to 5000. Initially total protein concentrations are set equally for each protein at $C_i = 0.1$ for $i = 1, 2, \cdots, 4$. Protein concentrations (determined *in vivo* by expression levels of corresponding genes and translation/degradation) values $C_i$ can fluctuate with rate determined by parameter $r$, (see below) unrelated to the mutation rate reflecting primarily the epigenetic factors such as long-time correlated extrinsic noise in protein production [26,54]. Due to the long time correlation in protein production levels, values of $C_i$ appear "inherited". Fluctuations in protein production levels are modeled in the following way. At each time step the value of $C_i$ may stay unchanged with probability $1-r$ or, with probability $r$, change. The magnitude of the change is random:

$$C_i^{new} = C_i^{old}(1+\varepsilon), \qquad (4)$$

where $C_i^{new}$ and $C_i^{old}$ are the new and old concentrations of protein product of $i$-th gene, $\varepsilon$ is drawn from Gaussian distribution whose mean and standard deviation are 0 and 0.1, respectively. Parameter $r$ characterizes the rate of fluctuations of protein copy numbers; we take $r = 0.01$ unless otherwise is noted.

The concentration of free (uncomplexed) proteins $F_i$ is determined from the LMA equations that assume that monomers and binary complexes can form:

$$F_i = \frac{C_i}{1 + \sum_{j=1}^{4} \frac{F_j}{K_{ij}}} \quad \text{for } i = 1, 2, \cdots, 4, \tag{5}$$

where $K_{ij}$ is the binding constant of interactions between protein $i$ and protein $j$ [55] and concentrations of binary complexes between all proteins (including homodimers) are given by the LMA relations:

$$F_{ij} = \frac{F_i F_j}{K_{ij}}. \tag{6}$$

We determined, after each change (a mutation or a fluctuation in $C_i$), all necessary quantities by solving the LMA Eqs. (5) and (6) to find $F_1$, $F_{23}$, and $F_{44}$ evaluate the new $P_{nat}$ for mutated protein(s) and $P_{int}^{23}$ and $P_{int}^{44}$ for the complex of protein pairs to be in their specific functional states as explained above. We solve coupled nonlinear LMA equations by iterations. Once $C_i$ changes or a mutation occurs, the old set of $F_i$ is substituted into the right hand side of Eq.(5) and a new set of $F_i$ is calculated. This procedure iterates until the difference between old and new values of $F_i$ drops below 0.1% of the new value.

To simulate a variable mutation rate, we consider protein 4 to be a component of DNA mismatch repair (MMR) machinery using an important part of it – *mut*S - as a prototype. Mutation rate at any time step t depends linearly on concentration of functional MMR dimers, which is

$$MMR_{funct}(t) = F_{44} P_{int}^{44} \left( P_{nat}^{4} \right)^2, \tag{7}$$

where the first term in the right hand side of Eq.(7) is a concentration of binary dimeric complex of the MMR protein, the second term is the thermal probability that homodimer forms a fixed functional conformation out of 144 possible docking modes and the last term stems from the requirement that both members of a functional MMR complex have to be in their native folded conformations.

The mutation rate depends linearly on concentration of functional MMR complexes:

$$m(t) = \begin{cases} m_0\left(1 - \dfrac{MMR_{funct}(t)}{MMR_{funct}(0)}\right) & \text{if } MMR_{funct}(t) < MMR_{funct}(0) \\ 0.0001 & \text{if } MMR_{funct}(t) \geq MMR_{funct}(0) \end{cases}, \qquad (8)$$

where $m_0 = 0.05$ is maximal mutation rate. We define as a mutator a clone whose mutation rate is greater than 0.01, *i.e.* 100 times or more higher than lowest (*i.e.* wild-type) value of 0.0001. The latter value is typical of non-mutator *E. coli* strains [34]. While the dependence presented by Eq.(8) is the most natural one, the results do not depend significantly on this assumption: a threshold-like dependence where mutation rates can take two values depending on whether $MMR_{funct}$ is below or above a certain threshold gives qualitatively the same adaptation behavior as presented here for the model Eq.(8) (data not shown)

In order to seed simulations with organisms whose initial state is non-mutator, we designed the initial sequences for protein 4 to be stable and to form a strong homodimer using design algorithms described in [41,48]. We did not initially design interactions between products of RCGs, so that populations start from non-adapted growth rate conditions.

*Acknowledgements.* We thank Louis Kang for his help at the initial stage of this work and Sergei Maslov for comments on the manuscript.

**Supplementary Figure**

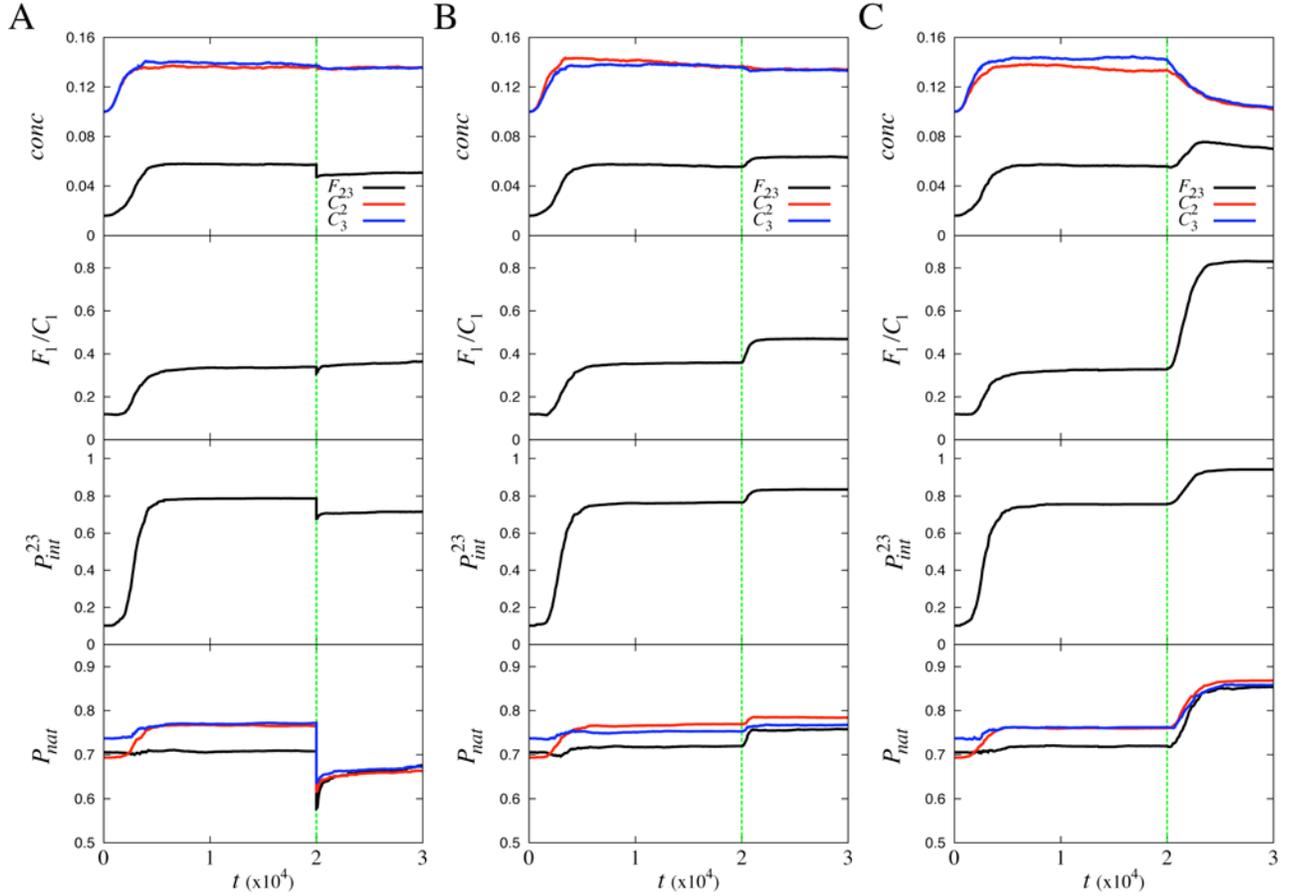

**Figure S1. Genotypic mutations: the main mechanism of adaptation.** In order to elucidate the main cause of increase in fitness upon adaptation, we investigated the time progression of microscopic physical-chemical quantities of RCGs, which determine fitness according to Eq.(1). We plot the concentration of complexes between protein 2 and 3 ($F_{23}$, black curve), the total concentrations of the protein 2 ($C_2$, red curve) and 3 ($C_3$, blue curve) in the top panels, the fractional concentration of protein 1 in its functional (monomeric) form ($F_1/C_1$) in the second panels, thermal probabilities to form functional complex between protein 2 and 3 ($P_{int}^{23}$) in the third panels, and stabilities of proteins 1 (black), 2 (red) and 3 (blue) ($P_{nat}$) in the bottom panels. The green lines at $t=20000$ marks the time of stress, which are (A) temperature increase, (B) 3 fold decrease of the base birth rate ($b_0$) (stationary phase), and (C) 10 fold drop of the optimal total concentration ($C_0$) of all proteins (starvation). The fractional concentrations of proteins are determined by the total concentration of proteins and the binding constants between all possible pairs of proteins in a cell according to the LMA equations as explained in the Main text. Changes in all of these quantities are due to genotypic mutations, which change properties of proteins such as their stability, solubility (affecting $F_1$ and $F_{23}$) and propensity for functional interaction $P_{int}^{23}$, independent on the phenotypic changes of the total protein concentrations.